# Multipole Orders in a Chain Ferrate $Na_2FeSe_2$

S. W. Lovesey[1,2] and D. D. Khalyavin[1]

[1]ISIS Facility, STFC, Didcot, Oxfordshire OX11 0QX, UK

[2]Diamond Light Source Ltd, Didcot, Oxfordshire OX11 0DE, UK

**Abstract** Fundamental block and staggered orders of magnetic Fe multipoles in $Na_2FeSe_2$ are classified by their symmetry and magnetoelectric properties. Our structure model incorporates ferromagnetic or antiferromagnetic coupling between chains. The ferrate salt is valued in studies of highly correlated electrons as the only iron selenide known to possess chain-like structural units hosting ferrous cations. Axial and polar (Dirac) multipoles are compulsory in the electronic structure since Fe ions exhibit enantiomorphic symmetry in the parent $K_2ZnO_2$-type compound. Calculated Bragg diffraction patterns for neutrons and x-rays reveal specific contributions from both multipole types.

*Introduction* – The study reported here is a contribution to an ongoing quest for a deeper understanding of electronic properties of magnetic materials with captivating properties. We make use of symmetry-based techniques capable of unveiling fundamental aspects of a material without approximations that are inevitable in specific calculations, e.g., application of one of the many band structure methods. Work of our kind is suited for compliance with Murray Gell-Mann's totalitarian principle, by which anything not forbidden (by symmetry) is compulsory. Specifically, parity-even and parity-odd magnetic multipoles that encapsulate electronic degrees of freedom in the ground state. The former (axial) multipole type is conventional magnetism, in the sense of its appearance in ichor-like haematite ($\alpha$-$Fe_2O_3$) and famed lodestone worried and written about from the time of Greek texts in 315 BC to William Gilbert of Colchester, the father of magnetism, in the 16th century, to Dzyaloshinskii in 1958 who gave a phenomenological theory of weak ferromagnetism. Parity-odd (polar) magnetism is a relatively new discovery, and intimately related to the property of electrically induced magnetization (magnetoelectric effect, ME) measured for the first time in 1960 using a sample of $Cr_2O_3$. In this Letter we classify axial and polar multipole orders by their magnetic symmetry and ME response, and predict their contribution to Bragg diffraction patterns available with beams of neutrons or x-rays. For the moment, there is scant knowledge about the chalcogenide considered apart from its crystal structure, because large samples are not yet available [1].

The chain ferrate salt $Na_2FeSe_2$ has aroused interest because it hosts ferrous cations and presents chain-like structural units [1, 2]. These properties are not shared by iron-based superconductors that have been shown to possess unusual magnetic orders [3, 4], or $TlFeS_2$ [5] and selenide semiconductor $TlFeSe_2$ [6] that appear to contain quasi-one-dimensional units. The two-leg ladder selenide $BaFe_2Se_3$ hosts ferrous ions, however, and it supports a complicated magnetic order that remains to be fully resolved [7, 8]. The variety of unusual magnetic phases presented by iron selenides is attributed to the interplay of spin, orbital and lattice degrees of freedom, and orbital-lattice mechanisms are effective for the atomic configuration $d^6$ with its single electron outside a closed shell.

Electrical and magnetic properties of haematite and chromium sesquioxide ($Cr_2O_3$) serve as orientation to properties of block and staggered states of $Na_2FeSe_2$. Haematite and $Cr_2O_3$ have the centrosymmetric corundum structure with Fe and Cr ions in enantiomorphic sites (3, $C_3$) and, likewise, the parent crystal structure of $Na_2FeSe_2$ is centrosymmetric with Fe ions in enantiomorphic sites. Furthermore, signatures of dipole moments along the trigonal axis are block and staggered for haematite and $Cr_2O_3$, respectively. The former is non-magnetoelectric (high-temperature modification, 260 K - 950 K, magnetic crystal-class 2/m), whereas $Cr_2O_3$ is a paradigm for the linear ME effect (magnetic crystal-class $\bar{3}'m'$).

$Na_2FeSe_2$ is composed of chain-like structural units in an I-centred orthorhombic $K_2ZnO_2$-type structure, with symmetry mmm ($D_{2h}$) [1]. Edge-sharing [$FeSe_4$] tetrahedra occupy the chemical structure Ibam (No. 72) with Fe (4a) at {(0, 0, 1/4), (0, 0, 3/4)}. Cations $Fe^{2+}$ are likely to have the high-spin $^5D$, J = 4, $3d^6$ atomic configuration. Atomic 3d states of a ferrous ion and the $K_2ZnO_2$-type crystal structure are nicely illustrated in Fig. 8 of Ref. [1].

There is growing conviction that $Na_2FeSe_2$ possesses magnetic states with signatures ↓↓↑↑↓↓ and ↓↑↓↑ for dipoles situated on a chain [1, 2]. Yet no mention to date of the inescapable fact that magnetic dipoles are conventional (axial) moments and anapoles. For ferrous ions ($Fe^{2+}$) in the parent compound occupy enantiomorphic sites (222 ($D_2$) point symmetry), and both axial and Dirac (polar) magnetic multipoles are present with the onset of magnetic order. Fundamental block and staggered states in the magnetic salt, with ferromagnetic or antiferromagnetic couplings between chains, are classified by a magnetic crystal-class and a ME property. Two of twelve possible motifs are depicted in Figs. 1 and 2. Calculated Bragg diffraction patterns prove that contributions from the multipole types are uniquely defined.

Interaction between neutrons and unpaired electrons is simple to formulate for small values of the scattering wave vector, **κ** [9]. Most authors analyse experimental data for elastic (Bragg) and inelastic scattering with an approximation to the interaction proportional to the magnetic moment, namely, {(2**S** + **L**) ⟨$j_0(κ)$⟩}, where **S** and **L** are operators for electronic spin and orbital angular momentum and the radial integral ⟨$j_0(κ)$⟩ is displayed in Fig. 3. An improved approximation {⟨**J**⟩ [g ⟨$j_0(κ)$⟩ + (2 − g) ⟨$j_2(κ)$⟩]}, where g the Landé splitting factor and the two radial integrals satisfy ⟨$j_0(0)$⟩ = 1 and ⟨$j_2(0)$⟩ = 0, is more appropriate for the analysis of Bragg diffraction patterns composed of many spots extending to large κ. Beyond dipoles, the quadrupole (K = 2) proportional to ⟨$j_2(κ)$⟩ is zero for states in a J-manifold, while the octupole (K = 3) for $3d^6$ has a form factor [⟨$j_2(κ)$⟩ − (4/3) ⟨$j_4(κ)$⟩]. It is compulsory to add Dirac multipoles to the foregoing axial multipoles when magnetic ions occupy acentric sites [9]. The Dirac dipole **D** contributes a term {i(**κ** × **D**)/κ} to the neutron-electron interaction. It is a sum of a spin anapole **Ω**$_S$ = **S** × **R**, orbital anapole **Ω**$_L$ = (**L** × **R** − **R** × **L**) and i**R**, where **R** is the operators for electronic position. Radial integrals that accompany each dipole are displayed in Fig. 3 [10,

11]. Likely, the Dirac dipole is significant in the range of wave vectors κ for which $\langle j_0(\kappa) \rangle$ is significantly different from zero. Dirac quadrupoles account for magnetic neutron diffraction by the pseudo-gap phases of ceramic superconductors YBCO and Hg1201[12].

*Models* - Multipole orders in $Na_2FeSe_2$ are derived from dipole moments parallel to crystal axes. Twelve orders are classified by a magnetic crystal-class and a ME property, which is a principal result of our study. (I) mm21′ with time-reversal 1′ permits a spontaneous dielectric polarization and a non-linear ME effect; (II) mmm′ with anti-inversion $\bar{1}$′ and a linear ME effect akin to the magnetic state of $Cr_2O_3$; (III) mmm1′ with all three inversions $\bar{1}$, 1′, $\bar{1}$′ and any kind of ME effect is prohibited. Orders (I) and (III) are antiferromagnetic motifs with nontrivial Bravais lattices defined by propagation vectors **k** = (0, 0, 1/2) and **k** = (1, 1, 1), respectively. A ferroelectric state allowed in (I) is realized in boracites, for example. Order (II) possesses a trivial antiferromagnetic motif with **k** = (0, 0, 0). All six block states ↓↓↑↑↓↓ under consideration belong to (I), while staggered states ↓↑↓↑ are divided between (II) or (III), the difference being antiferromagnetic (II) or ferromagnetic (III) coupling between chains. Magnetic space-groups for the multipole orders are gathered in Table I.

A structure factor for diffraction is $\Psi^K_Q = [\exp(i\boldsymbol{\kappa} \cdot \mathbf{d}) \langle O^K_Q \rangle_\mathbf{d}]$, where the Bragg wave vector $\boldsymbol{\kappa} = (h, k, l)$ with integer Miller indices, and the implied sum is over all Fe sites in a magnetic unit cell. A Hermitian electronic multipole $\langle O^K_Q \rangle$ of rank K with projections $-K \leq Q \leq K$ possesses discrete symmetries (Cartesian and spherical components of a dipole $\mathbf{R} = (x, y, z)$ are related by $x = (R_{-1} - R_{+1})/\sqrt{2}$, $y = i(R_{-1} + R_{+1})/\sqrt{2}$, $z = R_0$). In the case of magnetic diffraction, a multipole is time-odd ($\sigma_\theta = -1$) and parity-even ($\sigma_\pi = +1$, axial) or parity-odd ($\sigma_\pi = -1$, Dirac multipole).

*Bragg diffraction patterns* - First, we consider diffraction by six multipole orders in category (I). Following notation used in Table I, ferromagnetic (antiferromagnetic) coupling between chains is labelled α (β). Corresponding dipoles along the a-axis and b-axis are denoted by a shorthand $a_\alpha$ ($a_\beta$) and $b_\alpha$ ($b_\beta$). Four block states conforming to these dipole motifs all belong to space group $P_Cna2_1$ (No. 33.151 [14]), and $b_\beta$ is depicted in Fig. 1. The corresponding structure factor is,

$$\Psi^K_Q(33.151) \propto [1 + (-1)^l \sigma_\theta] [\langle O^K_Q \rangle + (-1)^{h+k} (-1)^K \sigma_\theta \sigma_\pi \langle O^K_{-Q} \rangle]. \quad (1)$$

Projections Q are odd, and $\langle O^K_{-Q} \rangle = (-1)^Q \langle O^K_Q \rangle^* = -\langle O^K_Q \rangle^*$. Base vectors (a, b, 2c) and (b, −a, 2c), with site symmetry $2'_c$, apply to block states using $a_\beta$ & $b_\alpha$ and $a_\alpha$ & $b_\beta$, respectively. Symmetry constraints mentioned apply to axial and Dirac multipoles, e.g., (1) applies to axial moments and anapoles on taking K = 1 and the appropriate values of time and parity signatures. The time signature is included in (1) since our electronic structure factors describes both Thomson scattering and resonance enhanced diffraction of x-rays [13], in addition to magnetic neutron diffraction of immediate interest ($\sigma_\theta = -1$). The selection rule *l* odd applies for magnetic diffraction, and it corresponds to a propagation vector **k** = (0, 0, 1/2). Referred to Miller indices for the parent structure ($H_o$, $K_o$, $L_o$) and base vectors (a, b, 2c), it follows that $h = H_o$, $k = K_o$, $l = 2L_o$ with $L_o$ half-integer. Bulk magnetism is forbidden, as expected. Diffraction amplitudes for axial dipoles and anapoles in the basal plane are 90° out of phase and corresponding intensities add. Moreover, the two dipoles are orthogonal and relative contributions are changed by the choice of Miller indices $h$ and $k$.

The structure factor (1) applies to the block state $c_\alpha$ described by $P_Cnn2$ (No. 34.161) with base vectors (a, b, 2c) and site symmetry $2_c$. Thus, projections Q are even integers. In consequence, dipoles (K = 1, Q = 0) obey strict selection rules $h + k$ even (axial) or $h + k$ odd (anapole). Selection rules for diffraction by the block state using $c_\beta$ are derived from $P_Cba2$ (No. 32.139). Base vectors and site symmetry for $c_\alpha$ & $c_\beta$ are the same. However, $\sigma_\theta$ does not occur as a coefficient of $\langle O^K_{-Q} \rangle$ in $\Psi^K_Q(32.139)$, which is otherwise the same as (1), and dipole selection rules are the reverse of those already mentioned for $\Psi^K_Q(34.161)$.

Turning to staggered multipole orders (II) and (III), the electronic structure factor,

$$\Psi^K_Q(57.392) \propto \langle O^K_Q \rangle [1 + (-1)^l \sigma_\theta \sigma_\pi] [1 + (-1)^{h+k+l} \sigma_\theta], \qquad (2)$$

with Q odd is appropriate for the staggered state using dipoles $a_\alpha$ and $b_\alpha$. Site symmetry $22'2'$ applies with base vectors (a, b, c) and (b, a, −c) for $a_\alpha$ and $b_\alpha$, respectively. I-centring is violated by multipole order (III) with propagation vector **k** = (1, 1, 1), and the selection rule on $l$ distinguishes between diffraction by axial ($\sigma_\theta \sigma_\pi = -1$) and Dirac ($\sigma_\theta \sigma_\pi = +1$) types. The latter feature is common to all staggered states under consideration. I-centring is restored in multipole order (II) using $a_\beta$ and $b_\beta$ described by $\Psi^K_Q(72.541)$, which is the same as (2) apart from replacement of the anti-translation selection rule by I-centring. Base vectors are (a, b, c) and (−b, a, c) with site symmetry $22'2'$ for $a_\beta$ and $b_\beta$, respectively, and a motif is depicted in Fig. 2. Staggered states using dipoles aligned with the c-axis have Q even from site symmetry $2'2'2$. Implementing this site symmetry, we find $\Psi^K_Q(56.376) = \Psi^K_Q(57.392)$ for $c_\alpha$ and base vectors (a, b, c). Finally, the staggered state using $c_\beta$ belongs to multipole order (II) described by $\Psi^K_Q(72.542)$ with base vectors (a, −b, −c).

Parity-even and parity-odd x-ray absorption events occur at different energies. Axial multipoles are observed in Bragg diffraction enhanced by an electric dipole - electric dipole (E1-E1) event, for example, while Dirac multipoles contribute to diffraction enhanced by an electric dipole - electric quadrupole event (E1-E2) [13]. Using the electronic structure factor (1), by way of an example, consider x-ray diffraction in the rotated channel of polarization (π′σ) and a Bragg wave vector (0, 0, $l$) with $l$ odd. In this case, diffraction enhanced by an E1-E1 event reveals the b-component of the axial dipole with an amplitude proportional to {cos(θ) cos(ψ)}, where θ is the Bragg angle and ψ the angle of rotation about the Bragg wave vector (azimuthal angle). By comparison, diffraction by an E1-E2 event reveals the a-component of the anapole and the amplitude is proportional to {sin(2θ) cos(ψ)}.

*Summary* - We enumerated motifs of magnetic multipoles in the ferrate salt $Na_2FeSe_2$. With chain-like structural units and ferrous cations it is a unique iron selenide material [1]. In terms of a standard cartoon using dipoles, our fundamental magnetic motifs are block states ↓↓↑↑↓↓ or staggered states ↓↑↓↑, with ferromagnetic (α) or antiferromagnetic (β) coupling between chains. Dipoles are conventional (axial) moments or anapoles (Dirac dipole). It is shown that motifs belong to one of three categories defined by a magnetic crystal-class and a magnetoelectric property. All magnetic space-groups are catalogued in Table I. Our electronic structure factors apply to multipoles of arbitrary rank, however, and they are used to predict Bragg diffraction patterns revealed by beams of neutrons or x-rays. An inelastic neutron scattering study of ferrous fluoride showed an unmissable hybridization of magnons and phonons [15, 16]. Vibrations in the electric crystal-field modulate the orbital state of the single electron outside the half-filled 3d-shell, and the mechanism might feature in magnetic excitations in $Na_2FeSe_2$. Likewise, $BaFe_2Se_3$ with two ladders of dipole moments. Space group $Pnm2_1$ (#31) has been used as a makeshift device to describe local structure in this material [7]. But if bulk polarization in $BaFe_2Se_3$ is zero, the parent structure is described by space group Pnma (not $Pnm2_1$), even if somewhere locally in the crystal there are some polar displacements. In fact, magnetic space-group $C_ac$ (No. 9.41, polar crystal-class m1') used by the authors of Ref. [8] follows if $Pnm2_1$ is substituted for the correct parent structure. The magnetic motif in $BaFe_2Se_3$ possesses a propagation vector **k** = (1/2, 1/2, 1/2) in the parent structure that is equivalent to $(1, 1, -1)_m$ in the monoclinic cell. The first magnetic Bragg spot $(1, 1, 0)_m$ equates to κ ≈ 0.71 Å$^{-1}$, or w ≈ 1.14 in Fig. 3 and all radial integrals in the anapole are seen to have significant values.

Acknowledgement. Professor G. van der Laan commented on the Letter in its making, and prepared Fig. 3 from calculations he performed.

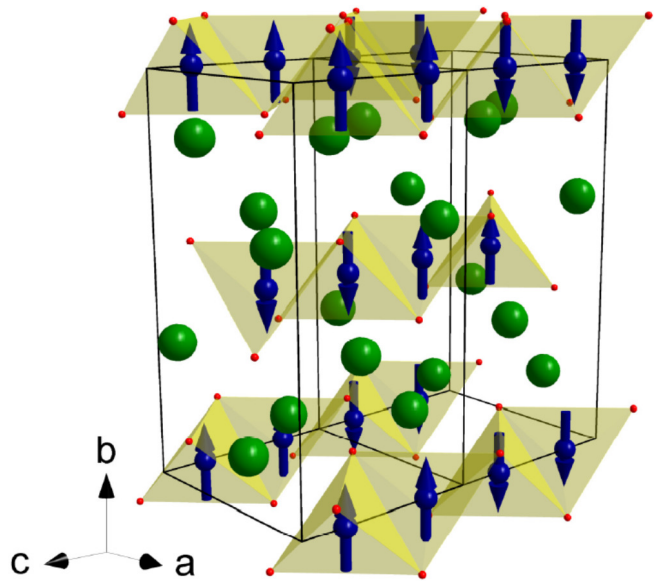

Fig. 1. Block state (I) in $Na_2FeSe_2$ with antiferromagnetic coupling between chains and dipole moments parallel to the b-axis (shorthand $b_\beta$). Magnetic space-group $P_Cna2_1$ (No. 33.151, BNS setting [14]), basis (b, −a, 2c) and propagation vector **k** = (0, 0, 1/2).

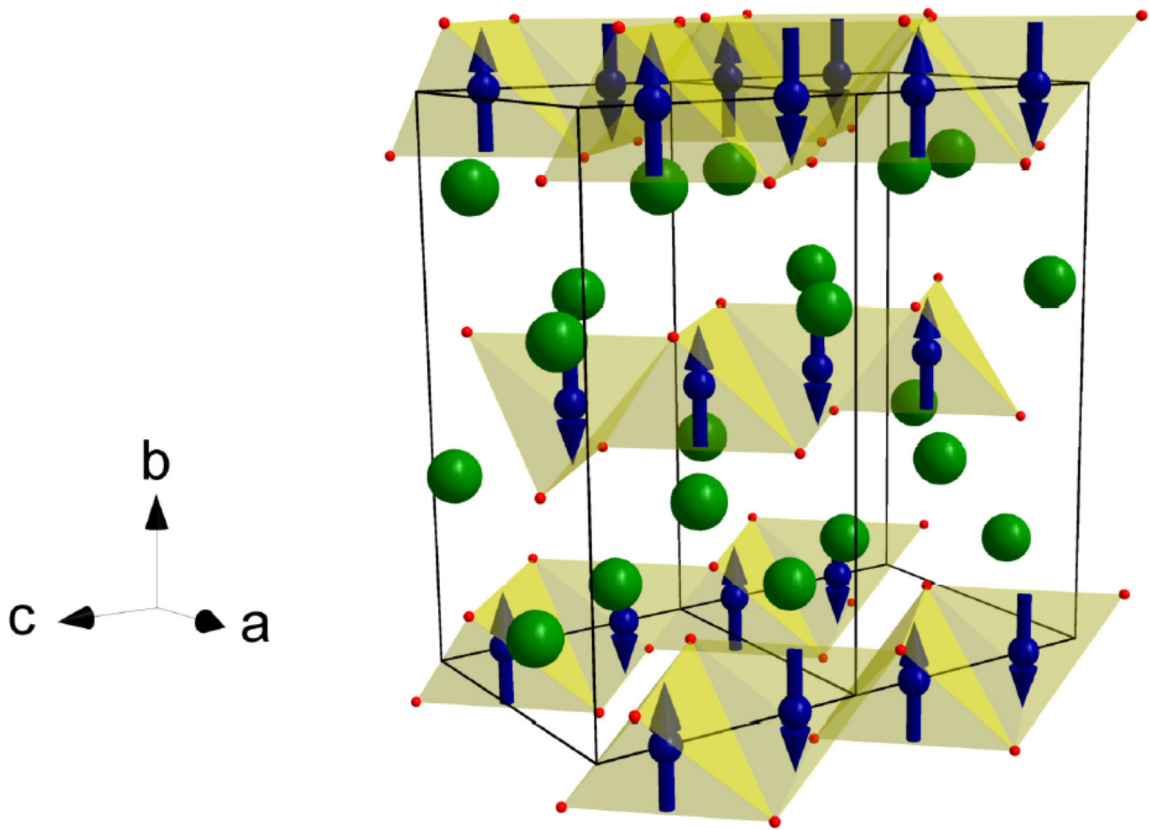

Fig. 2. Staggered state (II) with antiferromagnetic coupling between chains and dipole moments parallel to the b-axis (shorthand b$\beta$). Space-group Ib'am (No. 72.541), magnetic crystal class mmm', basis (−b, a, c) and propagation vector **k** = (0, 0, 0).

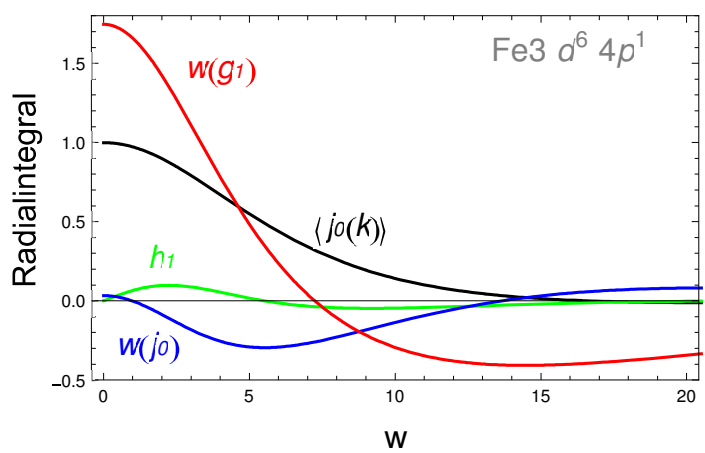

Fig. 3. Radial integrals for neutron diffraction by magnetic dipoles calculated using Cowan's atomic code [10, 11]. $\langle j_0(\kappa) \rangle$ in black for equivalent electrons $3d^6$ accompanies the magnetic moment. Atomic configuration Fe $(3d^6)$—Fe$(4p^1)$ is used for anapole radial integrals: (red) $\{w (g_1)\}$; (green) $(h_1)$; (blue) $\{w (j_0)\}$, where the dimensionless wave vector $w = 3a_o\kappa$ and $a_o$ is the Bohr radius. $\kappa = \{(4\pi/\lambda) \sin(\theta)\}$ where $\lambda$ and $\theta$ are the neutron wavelength and Bragg angle, respectively. Both $(g_1)$ and $(j_0)$ diverge in the limit $w \to 0$, and values displayed are multiplied by w. Integrals $(g_1)$, $(h_1)$, $(j_0)$ accompany dipoles i**R**, **Ω**$_S$, **Ω**$_L$, respectively, in the Dirac dipole, **D**.

TABLE I. Ferromagnetic and antiferromagnetic couplings between chains are labelled α and β, respectively. Magnetic space-groups are specified in BNS setting [14]. All staggered states use site symmetry 22′2′ for Fe ions with the dyad rotation operation 2 on the axis to which dipoles are aligned. Block states use site symmetry $2'_c$ for dipoles aligned along the a-axis and b-axis, and $2_c$ for alignment with the c-axis. Shorthand b$_\beta$ equates to dipoles aligned along the b-axis with antiferromagnetic coupling between chains, etc. Base vectors for each multipole order are listed.

(I) Block states: P$_C$nn2 (No. 34.161) c$_\alpha$ (a, b, 2c): P$_C$ba2 (No. 32.139) c$_\beta$ (a, b, 2c):
      P$_C$na2$_1$ (No. 33.151) a$_\alpha$ and b$_\beta$ (b, −a, 2c); a$_\beta$ and b$_\alpha$ (a, b, 2c)

(II) Staggered states: Ib′am (No. 72.541) a$_\beta$ (a, b, c); b$_\beta$ (−b, a, c):
      Ibam′ (No. 72.542) c$_\beta$ (a, −b, −c)

(III) Staggered states: P$_I$bcm (No. 57.392) a$_\alpha$ (a, b, c); b$_\alpha$ (b, a, −c):
      P$_I$ccn (No. 56.376) c$_\alpha$ (a, b, c)